\documentclass[11pt]{article}

\usepackage{scicite}

\usepackage{times}
\usepackage[latin1]{inputenc}
\usepackage{graphicx}
\usepackage{dcolumn}
\usepackage{bm}
\usepackage{bbding}
\usepackage{color}
\usepackage{amsmath}
\usepackage{mathptmx}
\usepackage{epstopdf}
\usepackage{ulem}
\usepackage[symbol]{footmisc}
\renewcommand{\thefootnote}{\fnsymbol{footnote}}

\definecolor{orange}{rgb}{1,0.5,0}
\definecolor{red}{rgb}{1,0,0}
  
\newcommand{\tcb}{\textcolor{black}}

\makeatletter
\def\@fnsymbol#1{\ensuremath{\ifcase#1\or 
\dagger\or \ddagger\or
   \mathsection\or \mathparagraph\or \|\or **\or \dagger\dagger
   \or \ddagger\ddagger \else\@ctrerr\fi}}
\makeatother

\newenvironment{sciabstract}{%
\begin{quote} \bf}
{\end{quote}}

\title{\tcb{Spin-down by dynamo action in simulated radiative stellar layers}}

\begin{document} 
\renewcommand{\thefootnote}{\fnsymbol{footnote}}

\author
{Ludovic Petitdemange$^{1\ast}$\thanks{These authors contributed equally to this work.}\, , Florence Marcotte$^{2\ast\dag}$, Christophe Gissinger$^{3,4\ast\dag}$\\
\\
\hspace{-5ex}\normalsize{$^{1}$ LERMA, Observatoire de Paris, PSL Research University, CNRS, Sorbonne Universit\'e, Paris, France}\\
\hspace{-5ex}\normalsize{$^{2}$ Universit\'e C\^ote d'Azur, Inria, CNRS, LJAD, Nice, France}\\
\hspace{-5ex}\normalsize{$^{3}$ Laboratoire de Physique de l'Ecole Normale Sup\'erieure, ENS, Universit\'e PSL, CNRS, Paris, France}\\
\hspace{-5ex}\normalsize{$^{4}$ Institut Universitaire de France, Paris, France}\\
\\
\hspace{-5ex}\normalsize{$^{\ast}$ Corresponding author. Emails:}\\ 
\hspace{-5ex}\normalsize{ludovic.petitdemange@upmc.fr, florence.marcotte@inria.fr, christophe.gissinger@phys.ens.fr.}\\
}


\date{}


\baselineskip24pt


\maketitle


\begin{sciabstract}
The evolution of a star is influenced by its internal rotation dynamics through transport and mixing mechanisms, which are poorly understood. Magnetic fields can play a role in transporting angular momentum and chemical elements, but the origin of magnetism in radiative stellar layers is unclear. Using global numerical simulations, we identify a subcritical transition \tcb{from laminar flow} to turbulence due to the generation of a magnetic dynamo. Our results have many of the properties of the theoretically-proposed Tayler-Spruit dynamo mechanism, which strongly enhances transport of angular momentum in radiative zones. It generates deep toroidal fields that are screened by the stellar outer layers. This mechanism could produce strong magnetic fields inside radiative stars, without an observable field on their surface.
\end{sciabstract}

As young stars form through accretion, or as ageing stars \tcb{have burned all of their hydrogen fuel}, the star's core contracts. Conservation of angular momentum causes \tcb{the core} to spin-up, producing strong gradients in the \tcb{angular velocity as a function of radius, a situation referred to as differential rotation}. Measurements using stellar pulsations (asteroseismology) have shown that \tcb{stars at various stages of their evolution} have internal rotation profiles that are flatter than expected from stellar evolution models, especially across \tcb{radiative zones, where outward transport of energy occurs through radiative diffusion rather than convection}~\cite{Ceillier2013, Deheuvels2014, VanReeth2018}. This discrepancy could be resolved if there is an unidentified mechanism that extracts angular momentum from the stellar core and suppresses differential rotation as the star evolves~\cite{Cantiello2014}.

A potential mechanism for \tcb{enhanced} angular momentum transport is stellar magnetism~\cite{Moss1992,Spruit1999,Braithwaite2017}. Magnetic fields on stellar surfaces can collimate jets of plasma~\cite{Albertazzi2014} or power flares~\cite{Fleishman2020}. Theoretical treatments of magnetic fields in radiative zone models have shown they modify the predicted dynamics of \tcb{stars}~\cite{Zahn1992,Maeder2003,Heger2004,Potter2012,Petit2017}. However, these predictions are limited by two theoretical problems. Firstly, magnetic fields are difficult to observe in deep stellar layers, including the radiative cores of \tcb{stars with less than $\sim 1.3$ solar mass. Even in stars where the radiative zone is located in the envelope, in most cases the amplitude of potential magnetic fields falls below the spectropolarimeters detection limit - except for 10\% of these stars where strong, dipolar magnetic fields have been measured~\cite{Wade2016}.} Secondly, the mechanism by which a dynamo magnetic field can be generated inside a radiative stellar layer remains unclear.

The dynamo instability is the spontaneous development of an amplification loop, by which \tcb{magnetic field directed toward the poles (poloidal field) is converted into magnetic field parallel to lines of latitudes (toroidal field) and vice versa}~\cite{Moffatt1978}. This conversion is mediated by plasma motions, which must be sufficiently powerful for \tcb{an initially weak magnetic field} to undergo self-amplification. In convective stellar layers, the required flow complexity can be provided by turbulent buoyant plumes~\cite{Charbonneau2014,Brun2017}. But in stably-stratified, radiative layers, dynamo action (and thus magnetic braking) require a different source of hydrodynamic turbulence. Several models have been considered to provide angular momentum transport, including internal waves~\cite{Rogers2013} or magnetic instabilities~\cite{Jouve2015}. Among the latter approach is the Tayler-Spruit (TS) dynamo model~\cite{Spruit2002,Maeder2004}. In this model, magnetic field generation in radiative layers relies on (i) the winding of poloidal field into toroidal one by differential rotation [the $\Omega$-effect \cite{Moffatt1978}], and (ii) the destabilization of the resulting strong, toroidal and axisymmetric magnetic field by the Tayler instability~\cite{Tayler1973}, which regenerates a poloidal field and thus (in theory) closes the dynamo loop initiated by differential rotation. However, global numerical simulations have not produced TS dynamos, casting doubt on whether the simplifications made in the theory \tcb{are valid when the plasma is turbulent}~\cite{Zahn2007}. It is therefore unclear whether a dynamo mechanism could operate in a stably-stratified stellar layer. We sought to numerically investigate whether a magnetic field can build up through dynamo instability, trigger magnetohydrodynamic turbulence and achieve efficient angular momentum transport in a radiative star.

We model a radiative stellar layer by considering the swirling flow of a stratified, non-ideal, electrically conducting fluid between two coaxial, spherical shells spinning at different rates. The intensity of the differential rotation is controlled by the dimensionless Rossby number ${Ro \equiv \Delta \varOmega/\varOmega}$, where $\varOmega$ and $\varOmega+\Delta \varOmega$ are the angular velocities of the outer and inner shell, respectively, and the strength of stratification is quantified by the buoyancy frequency $N$~\tcb{\cite{methods}}.

\begin{figure}
\includegraphics[width=\textwidth]{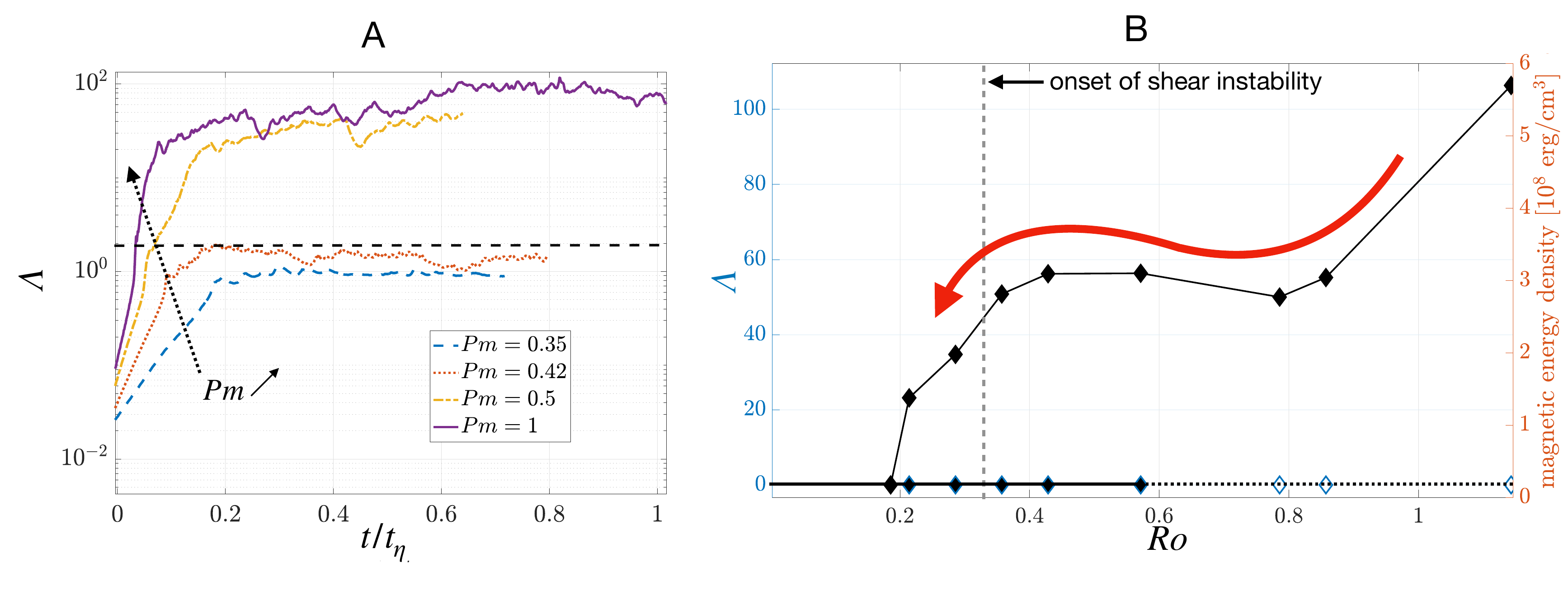}
\caption{{\bf Subcritical dynamo bifurcation.} (A)~Timeseries of the magnetic energy (measured by the \tcb{dimensionless} Elsasser number $\varLambda$) for \tcb{fixed rotation,  stratification and thermal diffusion} parameters \tcb{(specifically $E=10^{-5}$, $N/\varOmega=1.24$, $Pr=0.1$, $Ro=0.78$ as defined in \cite{methods})} and varying magnetic Prandtl number \tcb{($Pm=\{0.35\text{ (blue) }; 0.42\text{ (red) }; 0.5\text{ (yellow) }; 1\text{ (purple) }\}$, ordered as indicated by the dashed arrow)}. \tcb{Times ($t$) are made dimensionless using the Ohmic diffusion time $t_\eta$~\cite{methods}}. A secondary instability occurs when the magnetic energy exceeds a typical value $\varLambda_\circ \sim 1$ (dashed black line). (B)~Time-averaged magnetic energy density of the saturated dynamo as a function of shear rate (quantified by the Rossby number $Ro$), for $Pm=1$ (other parameters the same as panel A). Empty diamonds indicate linearly unstable solutions from which the magnetic field grows exponentially. Solid diamonds illustrate the \tcb{hysteresis cycle (black arrows)} between a non-magnetic solution \tcb{($\varLambda=0$)} and a strong, toroidal dynamo solution. \tcb{The scenario we propose follows the red arrow: the dynamo arises for initially large differential rotation, suppresses the shear as the star evolves and maintains the magnetic field below the stability threshold.}} 
  \label{fig:timeseries}
  \end{figure}

When the differential rotation across the radiative zone is weak \tcb{compared to overall rotation}, the flow is stable, axisymmetric and all velocity perturbations decay away rapidly. For steeper rotation profiles, the rotational invariance of the flow is broken by the destabilization of a free shear layer. In this case, Fig.~\ref{fig:timeseries}A shows the resulting magnetic energy. \tcb{Initially weak} magnetic fields undergo exponential amplification and saturate to a magnetic state, whose final amplitude can be tuned in our simulations by varying the fluid's ratio of molecular to magnetic diffusivities (the magnetic Prandtl number $Pm$). For large values of the plasma magnetic diffusivity ($Pm<0.5$), the shear instability amplifies a laminar and mostly axisymmetric toroidal dynamo which saturates at weak magnetic \tcb{energies (characterised by the dimensionless Elsasser number $\varLambda<1$,~\cite{methods})}. However, if the toroidal energy exceeds a transition value $\varLambda_\circ \sim 1$, a secondary instability is triggered: the exponential growth steepens, the system becomes turbulent, and the magnetic energy saturates at values nearly two orders of magnitude higher. We refer to these two dynamo solutions as the weak and strong dynamos, by analogy with Earth's dynamo~\cite{Roberts1978}. Fig.~\ref{fig:timeseries}B shows a bifurcation diagram that illustrates the subcritical behavior of the strong dynamo branch: once the dynamo has been \tcb{triggered} for steep rotation profiles, it can be maintained even when the shear rate is decreased well below the onset of hydrodynamic instability.

We propose that, \tcb{when stars exhaust their fuel}, the differential rotation across the radiative zone is large enough (high $Ro$) for extremely weak initial magnetic fields to be amplified by dynamo action. This induces enhanced outward transport of angular momentum, gradually flattening the rotation profile (decreasing the effective $Ro$) as the star evolves. The magnetic field then dynamically adjusts to the smoother rotation profile and sustains the turbulent motions on which it feeds, thus maintaining the magnetic field (Fig.~\ref{fig:timeseries}B).

\begin{figure}
\includegraphics[width=\textwidth]{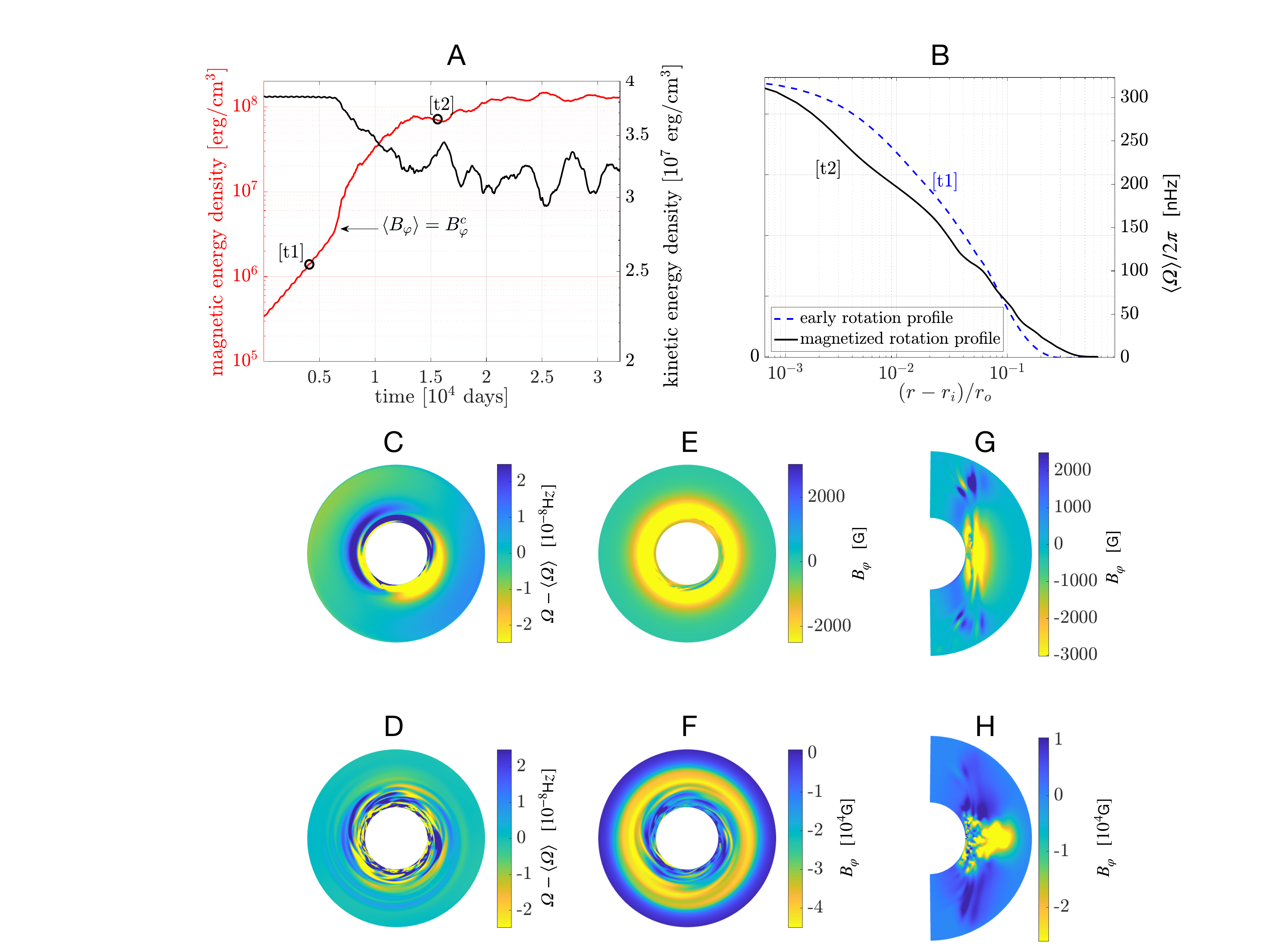}
\caption{ {\bf Transition to turbulence.} (A)~Timeseries of the total kinetic and magnetic energy densities. The arrow marks the time where the
amplitude of the axisymmetric component of the azimuthal magnetic field $\left \langle{B_\varphi}\right \rangle$ locally exceeds
the prediction of the Tayler instability, $B^C_\varphi$~\cite{Spruit2002}. \tcb{Angle brackets denote spatial averaging in the azimuthal direction, indicated by $\varphi$}. (B)~Radial profiles of the azimuthally-averaged angular velocity $\left \langle \varOmega \right \rangle$ in the equatorial plane for two distinct times, labelled [t1] and [t2] in the panel A, \tcb{with $r_i$ and $r_o$ the radii of the inner and outer shells, respectively.} The onset of the instability causes the rotation profile to flatten between [t1] and [t2]. Snapshots of (C and D) the non-axisymmetric angular velocity in the equatorial plane, (E and F) the azimuthal magnetic field in the equatorial plane and (G and H) the same quantity in the meridional plane.  Results are shown for the times [t1] (panels C, E and G) and [t2] (panels D, F and H), which are before and after the onset of secondary (Tayler) instability. This simulation has parameters: $E=10^{-5}$, $N/\varOmega=1.24$, $Pr=0.1$, $Ro=0.78$ and $Pm=1$.}
\label{fig:snapshots}
  \end{figure}

Fig.~\ref{fig:snapshots} shows snapshots of our simulations before and after the \tcb{steepening of the exponential growth} to illustrate how the strong dynamo causes the subcritical transition to hydrodynamic turbulence. \tcb{Steeper magnetic growth occurs at the same time as flow destabilization;} after this time, the magnetic field has become strongly chaotic, and exhibits fluctuations at small scale. The corresponding velocity field becomes highly turbulent, especially in the inner regions of the star. This transition to turbulent flow motions suppresses differential rotation across the fluid, causing flattening of the rotation (Fig.~\ref{fig:snapshots}B) as the strong dynamo builds up (Movie S1).

The strong dynamo that appears in our simulations shares several properties with the TS mechanism~\cite{Spruit2002}. Firstly, this dynamo feeds on the interaction between a large-scale, toroidal magnetic field and differential rotation. The resulting fields have a dominant axisymmetric, toroidal component, containing more than 80$\%$ of the magnetic energy. \tcb{Secondly, the spatial structure displayed by the magnetic field has a small lengthscale in the radial direction compared to the azimuthal direction.} Thirdly, our simulations show that strong dynamos arise when the maximum amplitude of the axisymmetric component of the azimuthal magnetic field 
exceeds the local stability threshold 
of the Tayler instability, in agreement with theoretical predictions~\cite{Spruit2002}. However, the TS dynamo loop in our simulations is initiated differently from the theoretical prediction~\cite{Spruit2002}, \tcb{which may explain why TS dynamos have long eluded numerical simulations (see Supplementary Text)}. In theoretical models, the finite-amplitude toroidal field required to trigger the Tayler instability is assumed to have grown (linearly) out of an \tcb{infinitesimal} poloidal field, wound up by differential rotation. In our simulations, this initial step is instead provided by the weak dynamo instability, which (exponentially) amplifies the toroidal field and kick-starts a subcritical TS dynamo. \tcb{Previous numerical simulations have shown that several mechanisms could play this role in stellar interiors (see Supplementary Text).} The dynamo transition we identify is turbulent, with many \tcb{unstable modes} excited simultaneously, which is consistent with laboratory analogs of the Tayler instability (see Supplementary Text). This indicates that a fluctuations-based TS dynamo occurs in our simulations, in which the axisymmetric field is replenished by the mean electromotive force, so magnetic fluctuations influence the saturation mechanism~\cite{Fuller2019}. The turbulent \tcb{nature of the} transition resolves a \tcb{controversy} in the predicted TS dynamo model, \tcb{that the dynamo loop cannot be closed with a single non-axisymmetric mode getting unstable at the onset of the Tayler instability, as winding up the latter would not alone replenish the required axisymmetric toroidal field}~\cite{Zahn2007} (see Supplementary Text).

\begin{figure}
\includegraphics[width=\textwidth]{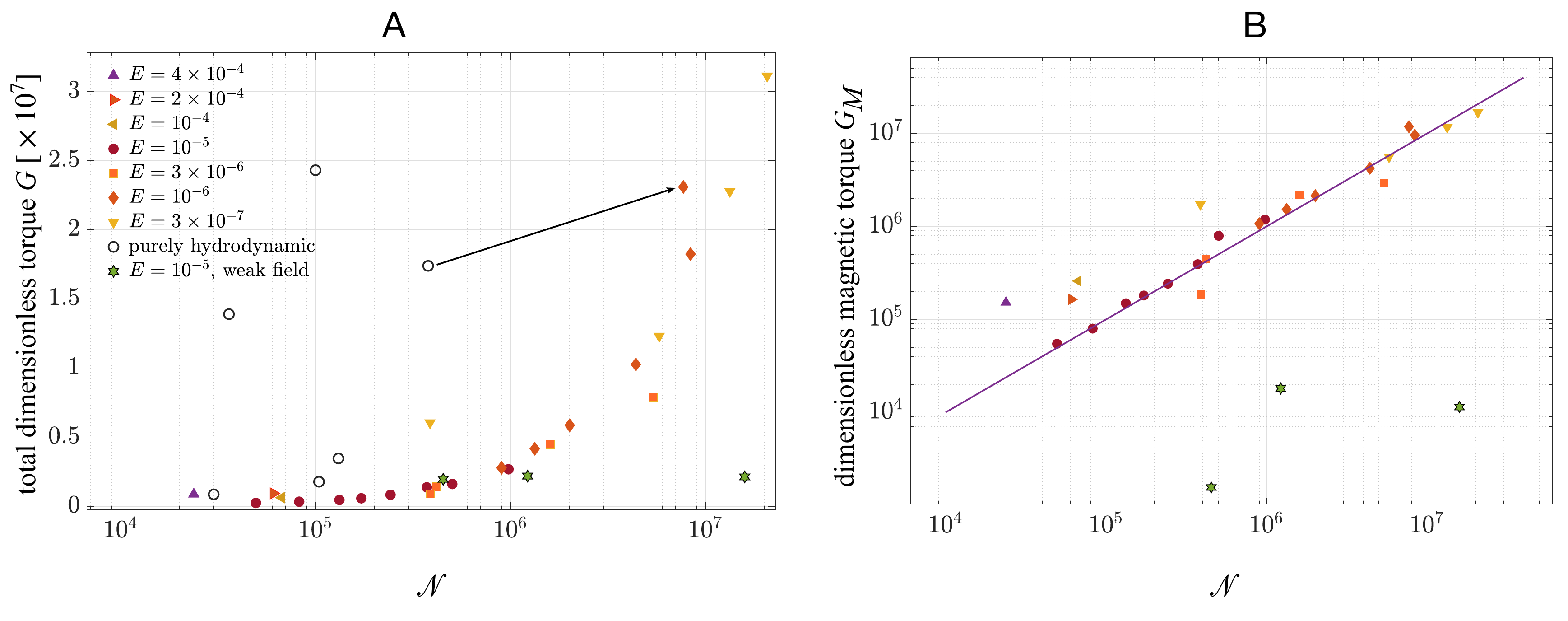}
\caption{{\bf Angular momentum transport.} (A)~Total dimensionless torque~\cite{methods} exerted on the inner sphere as a function of the \tcb{dimensionless quantity ${\cal N}\equiv \beta r_i^{5/2}(U_0\varOmega)^{3/2}/N\nu^2$~\cite{methods}}. Data points are for simulations that result in \tcb{non-magnetic flows} (empty circles), weak-dynamos (green \tcb{stars}), and strong-dynamos (red, purple, orange and yellow symbols, see legend). The arrow connects two simulations with identical parameters, with and without magnetic fields. (B)~Results for the magnetic torque only. The purple line is the theoretical prediction for the TS dynamo~\cite{Spruit2002}.}
  \label{fig:couple}
  \end{figure}

We quantify the enhanced transport of angular momentum to the outer regions of the star by measuring the total \tcb{($G$)} and magnetic \tcb{($G_{M}$)} torques exerted on the swirling fluid. We performed a range of simulations that systematically varied the overall rotation rate and the stratification of the radiative layers. \tcb{Fig.~\ref{fig:couple} shows the resulting torques for strong dynamo action, which follow a transposed version of the theoretically predicted powerlaw $G_{M} \sim {\cal N} \equiv \beta r_i^{5/2}\frac{(U_0\Omega)^{3/2}}{N\nu^2}$~\cite{Spruit2002}, where $r_i$ is the inner shell radius, $\nu$ the plasma viscosity, $U_0$ the local azimuthal velocity measured in the dynamo region and $\beta$ an adjustable parameter~\cite{methods}.} Weak-field dynamos do not follow this relation (Fig.~\ref{fig:couple}).

Our simulations produce a turbulent radiative dynamo that shares many features with the TS model. A strong magnetic field can be sustained by dynamo action inside a stably-stratified radiative zone, suppressing differential rotation and causing spin-down of the stellar core. As such, this dynamo provides a plausible mechanism to account for the \tcb{enhanced} transport of chemical elements and angular momentum \tcb{in non-convective stellar layers}. \tcb{In particular,} it also provides a potential additional transport mechanism for the radiative layers of solar-type stars \cite{Eggenberger2005}. Helioseismology has shown that the Sun has a flat rotation profile in its radiative zone~\cite{Howe2009}. 

The poloidal component of the dynamo mechanism we identify is extremely weak, so the resulting magnetic fields are almost entirely toroidal. They are also deep in the star's internal layers, where intense differential rotation takes place. The thick stellar outer layers screen them from the surface, preventing direct observations (but \tcb{they} could be inferred from asteroseismology~\cite{Mathis2021}). Our results therefore provide a physical mechanism \tcb{for enhanced transport of angular momentum in stellar interiors, through a dynamo action that produces no surface magnetic field.}\\

\large{\textbf{\textcolor{blue}{(Supplementary Materials appended below, p. 13--22)}}}
\normalsize

\pagebreak

\pagebreak

\pagebreak

\pagebreak

\clearpage

\bibliography{scibib}
\bibliographystyle{Science}

{\bf Acknowledgments:} This study used the HPC resources of MesoPSL financed by the R\'egion \^ile-de-France and the project EquipMeso (reference ANR-10-EQPX-29-01) of the programme Investissements d'Avenir supervised by the Agence Nationale pour la Recherche. Numerical simulations were also carried out at the CINES Occigen computing centers (GENCI project A001046698). {\bf Funding}: LP acknowledges financial support from ``Programme National de Physique Stellaire'' (PNPS) of CNRS/INSU, France. FM acknowledges financial support from the French program `T-ERC' managed by Agence Nationale de la Recherche (Grant ANR-19-ERC7-0008-01). CG acknowledges financial support from the French program `JCJC' managed by Agence Nationale de la Recherche (Grant ANR-19-CE30-0025-01).
{\bf  Competing interests}: The authors declare that they have no competing interests. {\bf Data and materials availability}: We used the {\sc Parody-JA} code \textit{(31,32)}, which is available at http://www.ipgp.fr/~aubert/tools.html, \tcb{combined with} the ShTns library \textit{(33)}, available at https://bitbucket.org/nschaeff/shtns/src/master/. \tcb{All the scripts, input parameters and output data required to reproduce the simulations  and evaluate the conclusions in the paper are available for download at https://doi.org/10.5281/zenodo.7419274 \textit{(34)}.}

\pagebreak

\begin{center}
\Large{\textbf{{Supplementary Materials}}}
\end{center}
Materials and Methods\\
Supplementary Text\\
Figures 4 to 7\\
Table S1\\
References (31-43)\\
Movie S1\\


\subsection*{ Materials and methods}

\noindent \underline{Model}

We consider an electrically conducting fluid, swirling between two concentric spheres of radius $r_i$ and $r_o$  that rotate about the same axis at different angular velocities ($\varOmega$ for the outer sphere and $\varOmega+\Delta \varOmega$ for the inner sphere), characterized by an aspect ratio $\chi=r_i/r_o$ set to $0.35$. Stable stratification is achieved inside the fluid by means of a prescribed temperature difference $\Delta T$ between the two shells. Using the Boussinesq approximation~\tcb{\textit{(35)}} to neglect variations in the fluid density except in the buoyancy term, the system is governed by the following magnetohydrodynamic equations, expressed in the reference frame where the outer sphere is at rest:
\begin{align}
\label{eq:full_set1}
\rho \left(\frac{\partial \bf{u}}{\partial t} \, + \, ({\bf u}\cdot{\bf \nabla}) {\bf u} \, + 2 \, \varOmega \, {\bf e_z}\times{\bf u} \right)&=   \, - {\bf \nabla} P \, + \, \frac{1}{\mu_0}{(\nabla \times \bf B)} \times{\bf B}  \,+   \,\alpha \rho g \varTheta \, {\bf e_r} \, + \,  \rho \nu \nabla^2 {\bf u},  \tag{S1}\\
\frac{\partial{\bf B}}{\partial t} \, - \, {\bf \nabla}\times({\bf u}\times{\bf B}) \, &= \,  \eta \, \nabla^2{\bf B}, \tag{S2} \\ 
\frac{\partial T}{\partial t} +({\bf u}\cdot{\bf \nabla}) T &= \kappa \, \nabla^2 T, \tag{S3}  \\
{\bf \nabla}\cdot{\bf u} &=0, \hspace{5mm}
{\bf \nabla}\cdot{\bf B}=0,\tag{S4a,b}
 \label{eq:full_set2}
\end{align}
where ${\bf u}$, $P$, ${\bf B}$ and $T$ are respectively the velocity, pressure, magnetic and temperature fields. $\varTheta$ is the temperature fluctuation accounting for density fluctuations in the buoyancy term, ${\bf e_z}$ the unit vector pointing along the rotation axis and ${\bf e_r}$ the local, radial unit vector. The fluid physical properties are described by the magnetic permeability $\mu_0$, the mean density $\rho$, the thermal expansion coefficient $\alpha$, the thermal diffusivity $\kappa$, the kinematic viscosity $\nu$ and the ohmic diffusivity $\eta$. 
The gravitational field, obtained by integrating the Poisson equation for a self-gravitating gas with spherical symmetry, is $g( r)\propto M( r)/r^2$ where $M( r)=\int_0^r \rho( r')4\pi r'^2dr'$ is the mass included in the sphere of radius $r$. In deep stellar interiors, the density $\rho$ is a slowly decreasing function of $r$, such that  $\rho$ can be considered constant and the gravitation field reduces to $g \propto r$~\textit{(36)}. The strength of the stratification is measured by the Brunt-V\"ais\"al\"a frequency (or buoyancy frequency) $N=(\alpha g_o\Delta T/(r_o-r_i))^{1/2}$, where \tcb{$g_o$ is the gravity} at the outer shell \tcb{$g(r_o)$}. Acoustic waves are filtered out by the use of Boussinesq approximation, which tends to stabilize the Tayler instability~\textit{(37)}. In the radiative zones we model, the difference of spinning rates $\Delta \varOmega$ between the inner and outer shells is fixed as a simulation parameter. This can be viewed as a way to focus only on a short period of the star's life. This restriction does not apply to a real radiative zone however, where the rotation profile could, eventually, flatten entirely over time.\\

\noindent \underline{Numerical setup}

All the simulations reported in this study were carried out using the {\sc Parody-JA} code~\textit{(31,32)} coupled with the ShTns library~\textit{(33)}.  {\sc Parody-JA} uses finite-difference discretization in the radial direction and spherical harmonics decomposition. \tcb{The number of radial gridpoints (${n_r}$) used in the fluid domain is $288<n_r<360$, and the maximal degree ($l_\text{max}$) and order ($m_\text{max}$) of the spherical harmonics decomposition are $128<l_\text{max}<188$ and $58<m_\text{max}<128$, respectively.} No-slip boundary conditions are applied on both spheres, along with electrically insulating boundary condition on the outer sphere, whereas the inner sphere has the same conductivity as the fluid. \\

\noindent \underline{Control parameters}

The flow regime is described by five independent, dimensionless parameters: the Ekman number $E = \nu / \varOmega r_o^{2}$ quantifying the ratio between the effects of viscous effect and Coriolis acceleration, the Rossby number $Ro=\Delta \varOmega/\varOmega$  comparing differential rotation and overall rotation rates, 
the Prandtl number $Pr = \nu / \kappa$ comparing molecular and thermal diffusivities, the magnetic Prandtl number $Pm=\nu/\eta$ comparing the kinematic and ohmic diffusivities, and the Rayleigh number $Ra = \alpha g_o\Delta T r_o^3 /( \nu\kappa)$ measuring the strength of the stratification, with $N/\varOmega= E\sqrt{Ra/Pr(1-\chi)}$. 
The thermal Prandtl number is fixed  to $Pr=0.1$. The hydrodynamic Reynolds number $Re = r_ir_o\Delta\varOmega / \nu$ measuring the ratio of inertial to viscous effects is obtained from the other dimensionless parameters as $Re=Ro\chi/E$. The magnitude of the total magnetic field is compared to Coriolis force through a  global Elsasser number calculated from the spatially averaged magnetic energy $\varLambda~=~\tfrac1V\int_V B^2/\rho \mu_0\varOmega \eta$, and a local \tcb{Elsasser number} $\varLambda_\text{local}=\overline{B_\varphi}^2/\mu_0\rho\varOmega\eta$ in which  $\overline{B_\varphi}$ is the maximum value of the axisymmetric component of the azimuthal field.

Although the numerical code is fully dimensionless, it can be described in terms of dimensional units. Apart from molecular diffusivities, which \tcb{due to overwhelming numerical cost} are orders of magnitude larger than those of real stars, our numerical simulations reproduce most of the typical parameters of stellar interiors: for example, the parameters of the simulation shown in Fig. 2 correspond to a stellar radiative zone of outer radius $r_o=7\times10^{8}$m, with global rotation rate of $\varOmega=3\times10^{-6}$s$^{-1}$;  density $ \rho=10^3$ kg m$^{-3}$;  buoyancy frequency $N=3.72\times10^{-6}$s$^{-1}$; shear strength $\Delta\varOmega=2.34\times10^{-6}$s$^{-1}$; magnetic diffusivity $\eta=1.5\times10^{7}$ m${^2}$s$^{-1}$; thermal diffusivity $\kappa=1.5\times10^{8} $m${^2}$s$^{-1}$ and kinematic viscosity $\nu=1.5\times10^{7}~$m${^2}$s$^{-1}$. More generally, using the same outer radius, density and global rotation rates, our simulations cover a range of buoyancy frequencies $N \in [10^{-6} \text{ to } 1.5\times10^{-4}]$s$^{-1}$ and shear strengths $\Delta\varOmega \in [10^{-7} \text{ to } 10^{-4}]$s$^{-1}$.\\

\noindent \underline{Typical timescales}

The Tayler instability occurs very fast, on an Alfvenic timescale (corresponding to a few turnover times only). Once generated, and despite considerable turbulent fluctuations, the dynamos are sustained for at least $10^3$ turnover times, the typical duration of our simulations. Fig.~S1 shows an extended energy timeseries for the same simulation as in Fig.~1 \& 2, integrated for more than $6$ ohmic times, where $t_\eta=(r_o-r_i)^2/\eta$ is the resistive timescale. The full time window in Fig.~4 is equivalent to $\sim 10^5$ turnover times, or $10^6$ days in dimensional units. These extended timeseries illustrate how fast the dynamo reaches a steady state. It also shows some hints of more complicated dynamics, involving a bistability between two TS states with slightly different energy levels. Failed dynamos \tcb{(corresponding to the simulations where no magnetic fields can be maintained)} typically experience exponential decay almost immediately, as expected in a fully turbulent flow in which the kinetic energy supply rate is related to the turnover time\tcb{~\textit{(16)}}. Our results therefore are unlikely to be a merely transient state. The resistive timescales in our simulations are extremely long compared to the turnover time, as they are in the Sun\tcb{~\textit{(17)}}.\\

\noindent \underline{Torques estimates}

The  amount of angular momentum that is transported to the outer regions of the star can be quantified by directly measuring the total azimuthal stress $S$ or the total torque $T=T_\nu+T_M=r_i^3S$  applied on the inner sphere, which includes the contribution of both the viscous torque $T_\nu$ and the magnetic torque $T_M$. 
We computed the corresponding dimensionless torque $G=T/(\rho\nu^2r_i)$ shown in Fig. 3 for a wide range of the parameters $E$, $Ra$, $Re$, with and without magnetic field. (For each simulation shown in Fig. 3, the estimated torque is time-averaged over the saturated stage.) The results show that the enhancement of the total torque between hydrodynamic runs and their MHD counterpart is limited by the imposed rotation rates of the \tcb{inner and outer spheres}, which yield artificial contributions to the viscous torque. To overcome this limitation, we compute the azimuthal stress $B_rB_\varphi/\mu$ exerted on the fluid by the dynamo field only. 
  
Fig.~5A shows two typical profiles of the magnetic field used to compute this torque. \tcb{These profiles} peak in the bulk \tcb{of the fluid} (\tcb{at radii} $r_0\sim 0.4-0.7$) and present a typical length scale $\lambda_{Ta}$ consistent with the theoretical prediction $\lambda_{Ta}\sim B_\varphi/\sqrt{\mu_0\rho}N$ \textit{(21)}. The magnetic torque $G_{M}$ is therefore computed in this region where the field is maximum. Fig.~5B shows that the velocity field always exhibits the same structure: a thin (Ekman) boundary layer develops at the inner sphere boundary, while the remaining velocity difference $\delta U_\varphi \sim U_0 =U_\varphi(r_0)$ is accommodated by the bulk shear flow over a typical scale $\lambda_{Ta}$. To compare our results to the theoretical prediction for the bulk of the flow, we therefore take the dimensionless differential rotation rate to be $q=kU_0/\varOmega$ with $k=2\pi/\lambda_{Ta}$ and systematically use $r_0=0.55$ for simplicity, \tcb{which also} avoids boundary layer effects.\\
 
 \noindent \underline{Scaling law for angular momentum transport}

Fig.~3B shows that TS-like dynamos scale as  $B_rB_\varphi/\mu\propto \rho(U_0\varOmega)^{3/2}/N$, or in dimensionless form:
\begin{equation}
{G_{M}}= {\cal N}\equiv \beta r_i^{5/2}\frac{(U_0\varOmega)^{3/2}}{N\nu^2},\tag{S5}
\label{scaling32}
\end{equation}
where $U_0$ is the local azimuthal velocity measured in the dynamo region and $\beta \sim 10^{-1}$ is an adjustable parameter standing for geometrical effects and for the effect of turbulent fluctuations on the value of the turbulent diffusivities. This scaling law corresponds to the theoretical prediction  $B_rB_\varphi/\mu\propto\rho\varOmega^2 r^2q^3 \left(\frac{\varOmega}{N} \right)^4$~\textit{(21)} in which the dimensionless differential rotation rate $q$ is expressed as $q\sim kU_0/\varOmega$, where $k=N\sqrt{\mu\rho}/B_\varphi$ is the radial wavenumber at which the Tayler instability takes place.

\subsection*{Supplementary text}

\tcb{Here we highlight several differences between the original Tayler-Spruit theory~\textit{(21)} and our simulations.}\\

\noindent \underline{Initiating the Tayler-Spruit loop}

The threshold of the Tayler instability is reached in our simulations through a primary, weak dynamo process driven by the shear instability, rather than by the winding up of the initial magnetic field (the $\Omega$-effect). This difference from the original Tayler-Spruit theory~\textit{(21)} \tcb{springs from the subcriticality of the Tayler instability: it} does not impact on the dynamo mechanism itself, but only on the way it is initiated. \tcb{The difference matters} because of the numerical difficulty in achieving a parameter regime where the $\Omega$-effect is sufficiently vigorous to meet the required toroidal field amplitude for the Tayler instability. \tcb{This numerical difficulty may explain why some previous attempts to produce TS dynamos were unsuccessful. In our simulations, the issue of initiating the TS dynamo is overcome because a weak dynamo process could be obtained at affordable numerical cost in the explored parameter regime, producing the toroidal field required to trigger the Tayler instability and  kickstart the TS dynamo.} Several dynamo mechanisms could provide this primary amplification in real stars~\textit{(38-41)}. In our simulations, the Tayler instability is triggered when the amplitude of the axisymmetric, toroidal field reaches the value predicted for dissipative systems~\textit{(21)}, $\varLambda_\text{local}>\sqrt{\frac{RaE\chi^2}{1-\chi}}$ (shown in Fig.~1~and~2).\\

\noindent \underline{Subcritical transition to turbulence \tcb{and mean-field dynamo}}

Once the magnetic field becomes unstable to the Tayler instability, the turbulent state generated by the subcritical dynamo differs from the original Tayler-Spruit scenario. Fig.~6 shows power spectra \tcb{density} of the velocity and magnetic fields for a typical simulation, illustrating how many azimuthal modes are excited simultaneously, similarly to the behavior observed in experimental studies of the Tayler instability~\textit{(42)}. The magnetostrophic force balance produces similar spectra for velocity and magnetic fields, and the energy displays a decreasing continuous spectrum of $m>0$ modes, with a quadrupolar symmetry at large scale. This spectrum indicates a mean-field dynamo in which the electromotive force \tcb{$\left \langle {\bf u \times \bf B} \right \rangle$} due to non-axisymmetric modes generates axisymmetric large scale magnetic fields~\textit{(43)}, with a $m=0$ mode peaking one order of magnitude above the non-axisymmetric modes. This subcritical transition to turbulence bypasses a criticism of the original Tayler-Spruit mechanism~\textit{(24)}, that the ``pure'' $m=1$ non-axisymmetric field produced by Tayler instability near its onset would not be sufficient to regenerate the $m=0$ toroidal field required to maintain the instability. The subcriticality of the transition makes it difficult to obtain numerically and track in the parameter space.\\

\noindent \underline{Stratification effects}

The theoretical derivation of the TS dynamo~\textit{(21)} relies on the assumption that $N \gg \varOmega$ (which allows for the so-called shellular approximation\tcb{~\textit{(10)}, that the angular velocity is invariant on each shell and depends only on the radial coordinate}). This is not the case for the dynamos in our simulations (even though a minimal stratification seems to be required, because no strong, TS-like dynamos were found for $N/\varOmega<0.1$). These strong dynamos were observed over a wide range of differential rotation and stratification profiles, spanning almost one order of magnitude in $Ro$ and showing no sign of inhibition at large values of the stratification $N/\varOmega \gg 1$. Fig.~7 shows snapshots of a dynamo obtained for $N/\varOmega=50$ and integrated for roughly $10^4$ turnover times. The magnetic field is similar to \tcb{that shown in Fig.~2} for smaller values of $N/\varOmega$, with a strong toroidal magnetic field in the equatorial plane associated with multiple non-axisymmetric modes and turbulent features. However, both magnetic and velocity fields have a spherical rather than cylindrical geometry, as expected in stars where density stratification dominates rotation\tcb{~\textit{(10)}}.

The  torque associated with this strongly stratified TS dynamo is large and similar to those obtained at lower values of $N/\Omega$. However, due to  numerical limitations, this regime $N/\Omega\gg1$ could only be reached in the less realistic limit $Ro\gg1$.\\


\pagebreak

\begin{figure}[h]
    \centering
\includegraphics[width=0.6\textwidth]{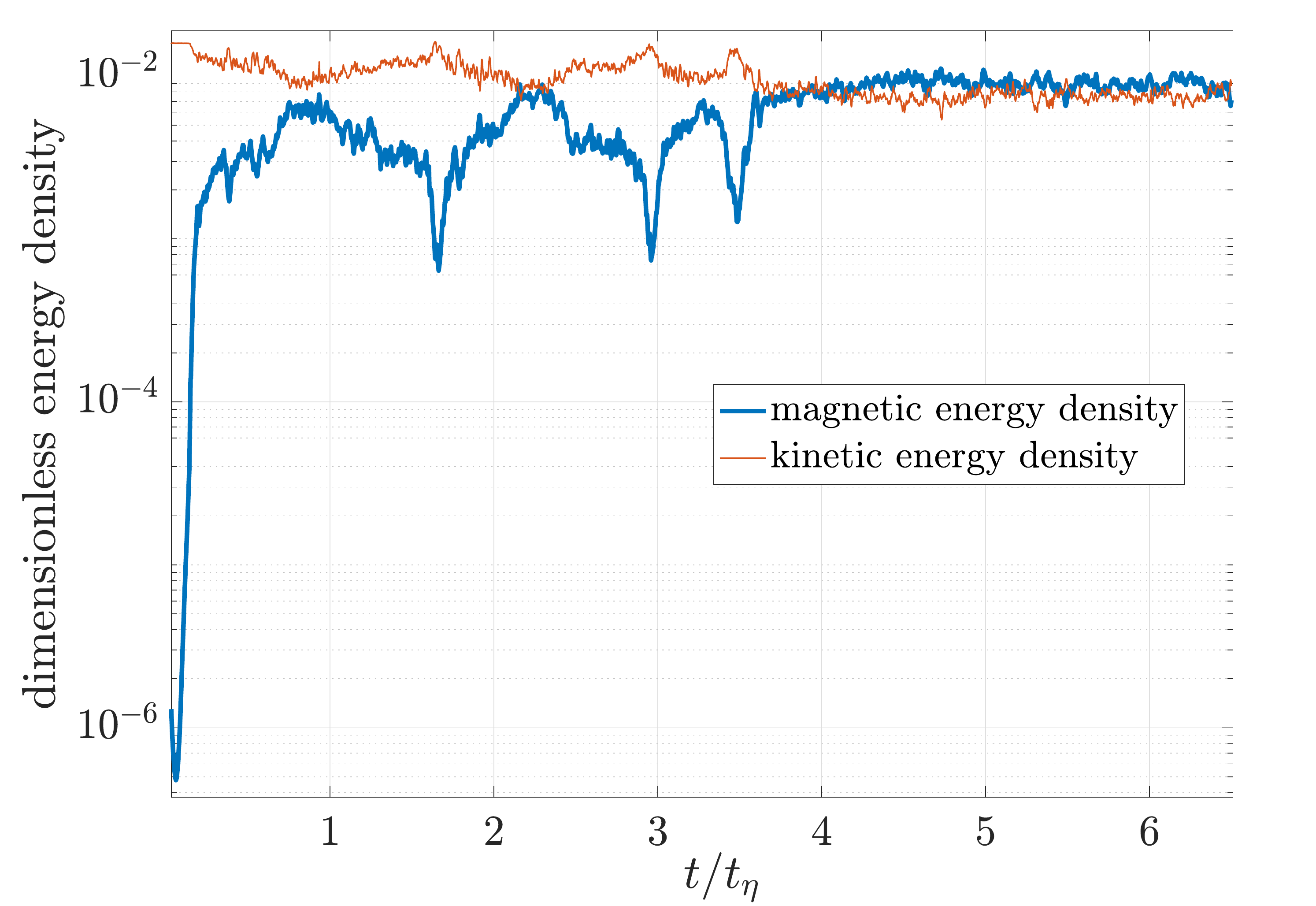}
     \caption{{\bf Extended energy timeseries}. The simulation was integrated over several ohmic times, where $t_\eta=(r_o-r_i)^2/\eta$. \tcb{Both magnetic and kinetic energies} are normalized by the typical kinetic energy $\rho \Delta\varOmega^2r_i^2$, to illustrate the magnetostrophic force balance reached by the dynamo by the end of the timeseries.} 
     \label{fig:run_ref_long}
\end{figure}

\pagebreak

  \begin{figure}[h]
\includegraphics[width=\textwidth]{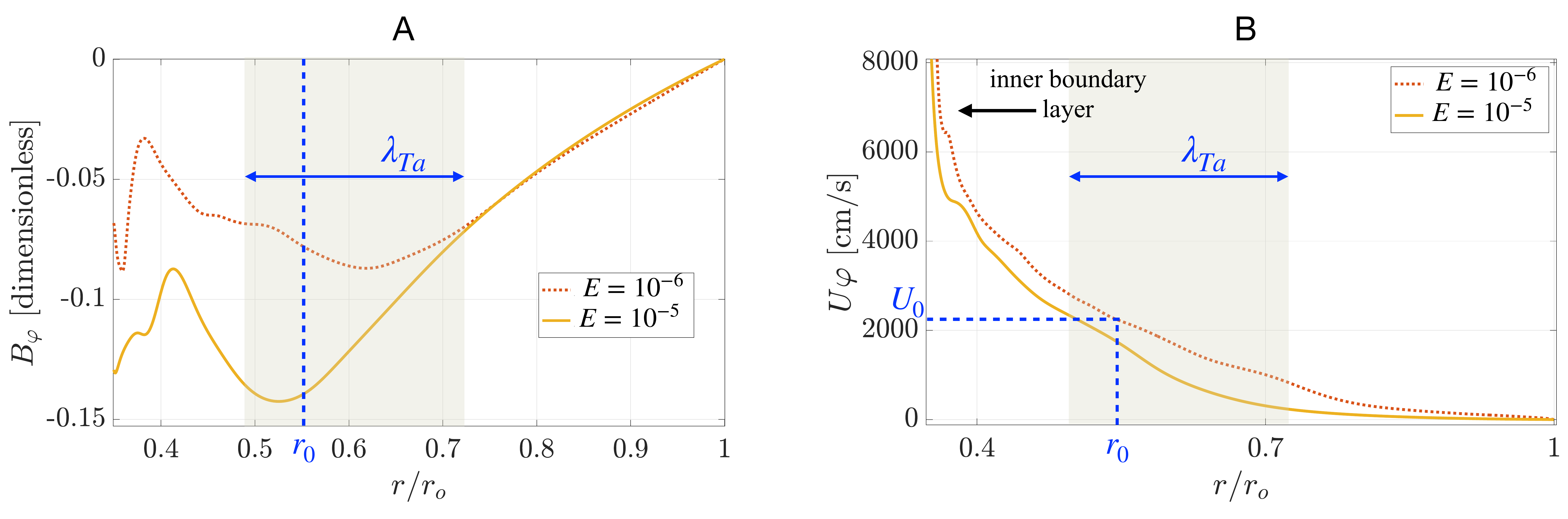}
\caption{{\bf Strong dynamo branch: flow profiles.}  (A) Radial profiles of the (azimuthally averaged) azimuthal magnetic field in the equatorial plane for two examples of TS-like dynamos at steady state ($E=10^{-5}$, $Re=10^4$, $Ra=10^9$ in yellow and $E=10^{-6}$, $Re=3 \times 10^4$, $Ra~=~10^{10}$ in red). (B) Radial profiles of the (azimuthally averaged) azimuthal velocity field in the equatorial plane, for the same simulations. \tcb{The blue label $r_0$ denotes the radius at which magnetic torques are systematically computed in Fig.~3, and $U_0$ corresponds to the (azimuthally averaged) azimuthal velocity at radius $r_0$, which we use to evaluate the right-hand side of (\ref{scaling32}).} \tcb{The shaded area marks the region of the domain, of typical width $\lambda_{Ta}$, where TS dynamo action takes place. The black arrow signals the presence of a viscous boundary layer, as shown by the sharp velocity gradient near the inner boundary.}}
\label{fig:profils}
  \end{figure}

\pagebreak

\begin{figure}[h]
    \centering
\includegraphics[width=\textwidth]{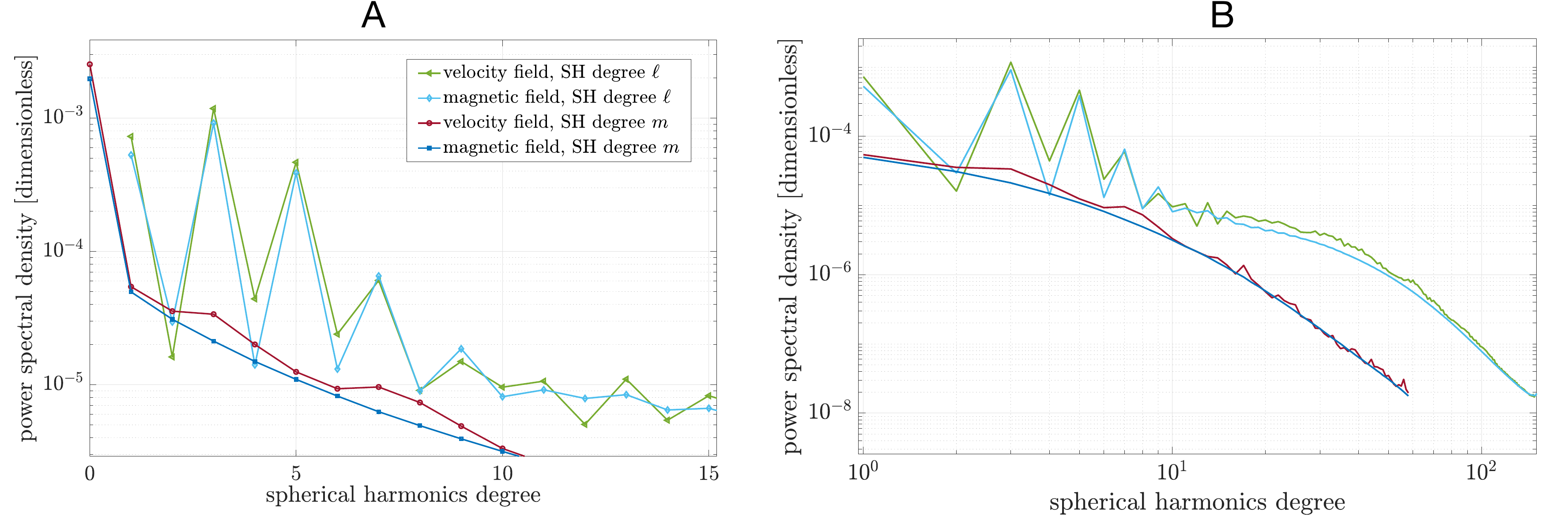}
\caption{{\bf Spectral densities.} (A) Power spectra of the velocity and magnetic fields of the first spherical harmonic degrees $m$ and $\ell$. (B) The same data over a wider range of spherical harmonics. The simulation parameters are $E=10^{-5}$, $Re=2.75\times10^{4}$, $Pm=1$, $Pr=10^{-1}$ and $Ra=10^{9}$.}
    \label{fig:spectrum}
\end{figure}

\pagebreak

  \begin{figure}[h]
    \centering
\includegraphics[width=\textwidth]{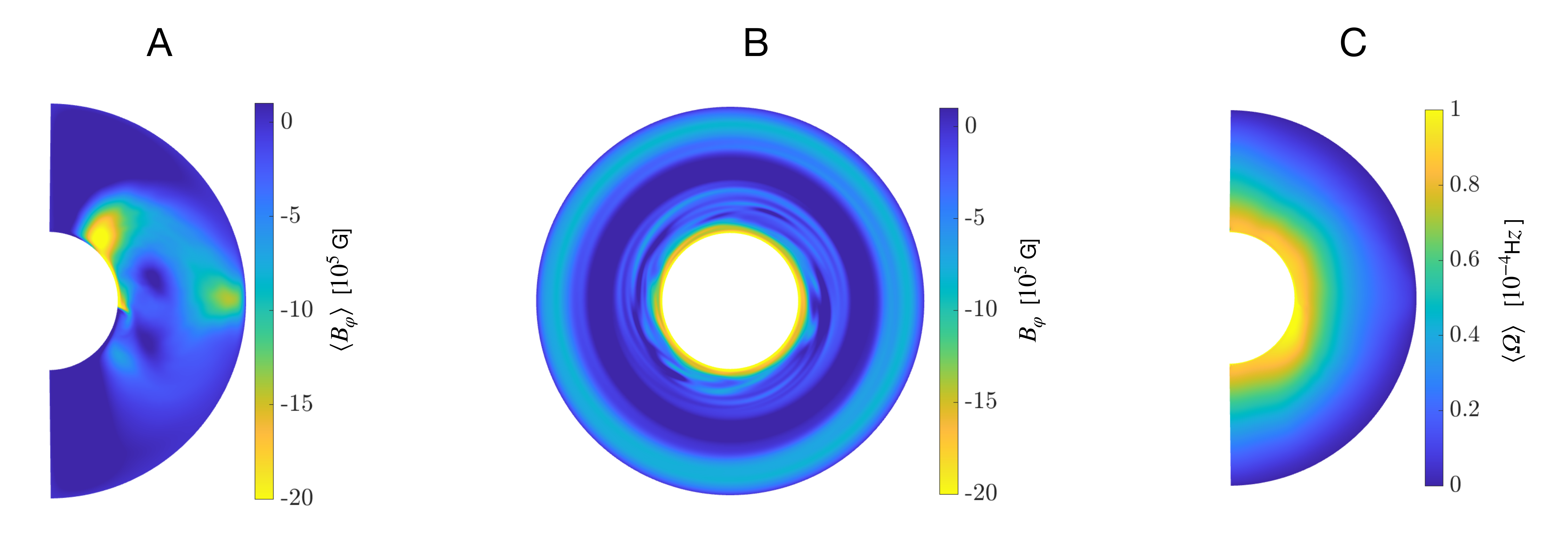}
     \caption{{\bf Strongly stratified regime}. Snapshots of (A) the azimuthal magnetic field in the meridional section, averaged in the azimuthal direction and over the steady state, (B) instantaneous azimuthal magnetic field in the equatorial plane, and (C) angular velocity in a meridional section, averaged in the azimuthal direction and over the steady state.
 This simulation has $N/\varOmega\approx50$ ($E=4\times10^{-4}$, $Ra=10^9$, $Re=4\times10^{4}$, $Pr=10^{-1}$, $Pm=1$).}
    \label{fig:N50}
\end{figure}

\pagebreak

{\bf Caption for Movie S1 (separate file). \tcb{Strong dynamo and transition to turbulence}.} (A)~Time series of the total kinetic and magnetic energy densities. The ratio $\left \langle {B_\varphi} \right \rangle /B^c_\varphi$ marks the time where the amplitude of the axisymmetric component of the azimuthal magnetic field locally exceeds the threshold for Tayler instability ($\left \langle{B_\varphi}\right \rangle /B^c_\varphi=1$), in agreement with the theoretical prediction for these parameter values~{(\textit{21})}. Snapshots of (B)~the azimuthal magnetic field in a meridional plane, (C)~the non-axisymmetric angular velocity in the equatorial plane and (D)~the azimuthal magnetic field in the equatorial plane. (E)~Animated radial profiles of the azimuthally-averaged angular velocity in the equatorial plane.  As the threshold for the Tayler instability is reached (see panel A), both magnetic and velocity fields become turbulent (see panels B,C,D) and the resulting enhanced transport of angular momentum causes a flattening of the rotation profile (see panel E). This transition to turbulence is associated with amplification of the magnetic energy by nearly two orders of magnitude (see panel A). Simulation parameters: $E = 10^{-5}$, $N/\varOmega = 1.24$, $Pr = 0.1$, $Ro = 0.78$ and $Pm = 1$.

\pagebreak

\begin{table}[h]
\centering
\vspace{-1cm}
\caption{{\bf List of simulation runs}. Input dimensionless parameters are listed for each simulation we ran. The last column indicates whether the run produced a strong or weak dynamo, or was a purely hydrodynamic simulation.}
\centering
\begin{tabular}{ l l  l  l  l  l l l}
\hline
                           $E$  & $Ra$ & $Re$ & $Pr$ & $Pm$ & $N/\varOmega$ & duration of the run [days] & Type  \\
\hline
 $4\times10^{-4}$ &  $10^9$   & $4\times10^4$   & $10^{-1}$  & 1.0 & $49.6$   & $4.0775\times 10^3$ &Strong   \\
 $2\times10^{-4}$ &  $10^9$   & $4\times10^4$   & $10^{-1}$  & 1.0 & $24.8$   & $5.3398\times 10^3$ &Strong   \\
 $10^{-4}$ &  $10^8$   & $2\times10^4$   & $10^{-1}$  & 1.0 & $3.9$   & $3.6042\times 10^3$ & Strong   \\

 $10^{-5}$ &  $10^9$   & $4\times10^4$   & $10^{-1}$  & 1.0 & $1.24$   & $7.5547\times 10^4$ &Strong   \\
 $10^{-5}$ &  $10^9$   & $3\times10^4$   & $10^{-1}$  & 1.0 & $1.24$   & $1.0194\times 10^5$ &Strong   \\
 $10^{-5}$ &  $10^9$   & $2.75\times10^4$   & $10^{-1}$  & 1.0 & $1.24$   & $1.0582\times 10^6$ &Strong   \\
 $10^{-5}$ &  $10^9$   & $2\times10^4$   & $10^{-1}$  & 1.0 & $1.24$   & $1.4437\times 10^5$ &Strong   \\
 $10^{-5}$ &  $10^9$   & $1.5\times10^4$   & $10^{-1}$  & 1.0 & $1.24$   & $9.6736\times 10^4$ &Strong   \\
 $10^{-5}$ &  $10^9$   & $1.25\times10^4$   & $10^{-1}$  & 1.0 & $1.24$   & $4.6000\times 10^4$ &Strong   \\
 $10^{-5}$ &  $10^9$   & $10^4$   & $10^{-1}$  & 1.0 & $1.24$   & $5.0417\times 10^4$ &Strong   \\
 $10^{-5}$ &  $10^9$   & $7.5\times10^3$   & $10^{-1}$  & 1.0 & $1.24$   & $1.9319\times 10^5$ &Strong   \\
 $10^{-5}$ &  $10^9$   & $2.75\times10^4$   & $10^{-1}$  & 0.5 & $1.24$   & $6.4242\times 10^4$ &Strong   \\

 $3\times10^{-6}$ &  $10^{10}$   & $9\times10^4$   & $10^{-1}$  & 1.0 & $1.18$   & $2.4473\times 10^4$ &Strong   \\
 $3\times10^{-6}$ &  $10^{10}$   & $6\times10^4$   & $10^{-1}$  & 1.0 & $1.18$   & $2.2975\times 10^4$ &Strong   \\
 $3\times10^{-6}$ &  $10^{10}$   & $3\times10^4$   & $10^{-1}$  & 1.0 & $1.18$   & $4.2978\times 10^4$ &Strong   \\
 $3\times10^{-6}$ &  $10^{10}$   & $2\times10^4$   & $10^{-1}$  & 1.0 & $1.18$   & $1.0764\times 10^5$ &Strong   \\

 $10^{-6}$ &  $10^{10}$   & $12\times10^4$   & $10^{-1}$  & 1.0 & $0.39$   & $4.1956\times 10^3$ &Strong   \\
 $10^{-6}$ &  $10^{10}$   & $9\times10^4$   & $10^{-1}$  & 1.0 & $0.39$   & $3.0247\times 10^4$ &Strong   \\
 $10^{-6}$ &  $10^{10}$   & $6\times10^4$   & $10^{-1}$  & 1.0 & $0.39$   & $3.7037\times 10^4$ &Strong   \\
 $10^{-6}$ &  $10^{10}$   & $4\times10^4$   & $10^{-1}$  & 1.0 & $0.39$   & $3.4549\times 10^4$ &Strong   \\
 $10^{-6}$ &  $10^{10}$   & $3\times10^4$   & $10^{-1}$  & 1.0 & $0.39$   & $6.8519\times 10^4$ &Strong   \\
 $10^{-6}$ &  $10^{10}$   & $2\times10^4$   & $10^{-1}$  & 1.0 & $0.39$   & $1.5833\times 10^5$ &Strong   \\

 $3\times10^{-7}$ &  $10^{10}$   & $8\times10^4$   & $10^{-1}$  & 1.0 & $0.12$   & $1.7699\times 10^4$ &Strong   \\
 $3\times10^{-7}$ &  $10^{10}$   & $6\times10^4$   & $10^{-1}$  & 1.0 & $0.12$   & $4.7775\times 10^4$ &Strong   \\
 $3\times10^{-7}$ &  $10^{10}$   & $4\times10^4$   & $10^{-1}$  & 1.0 & $0.12$   & $2.6331\times 10^4$ &Strong   \\
 $3\times10^{-7}$ &  $10^{10}$   & $2\times10^4$   & $10^{-1}$  & 1.0 & $0.12$   & $9.8187\times 10^4$ &Strong   \\

 $10^{-5}$ &  $10^4$   & $3\times10^4$   & $10^{-1}$  & 1.0 & $4\times10^{-3}$   & $5.9259\times 10^3$ &Weak   \\
 $10^{-5}$ &  $10^6$   & $3\times10^4$   & $10^{-1}$  & 1.0 & $4\times10^{-2}$   & $2.9861\times 10^3$ &Weak   \\
 $10^{-5}$ &  $10^8$   & $3\times10^4$   & $10^{-1}$  & 1.0 & $0.4$   & $3.0093\times 10^3$ &Weak   \\
 $10^{-5}$ &  $10^9$   & $2.75\times10^4$   & $10^{-1}$  & 0.35 & $1.24$   &$6.3636\times 10^4$ & Weak   \\
 $10^{-5}$ &  $10^9$   & $2.75\times10^4$   & $10^{-1}$  & 0.42 & $1.24$   & $7.0189\times 10^4$ &Weak   \\

 $10^{-5}$ &  $10^9$   & $4\times10^4$   & $10^{-1}$  & N/A & $1.24$   & $2.0833\times 10^3$ &Hydro   \\
 $10^{-5}$ &  $10^9$   & $2.75\times10^4$   & $10^{-1}$  & N/A & $1.24$   &  $1.2462\times 10^4$ &Hydro   \\
 $10^{-6}$ &  $10^{10}$   & $4\times10^4$   & $10^{-1}$  & N/A & $0.39$   &  $1.1823\times 10^5$ &Hydro  \\
 $10^{-6}$ &  $10^{10}$   & $12\times10^4$   & $10^{-1}$  & N/A & $0.39$   & $1.5770\times 10^4$ &Hydro   \\
 $3\times10^{-7}$ &  $10^{10}$   & $6\times10^4$   & $10^{-1}$  & N/A & $0.12$   &  $8.2305\times 10^3$ &Hydro   \\
 $10^{-7}$ &  $10^{10}$   & $6\times10^4$   & $10^{-1}$  & N/A & $3.9\times10^{-2}$   & $3.7230\times 10^4$ & Hydro  \\
   &    &    &    &    &  &    &       \\
   &    &    &    &    &   &  &      
\end{tabular}
\end{table}

\end{document}